\documentclass[
superscriptaddress,
amsmath,amssymb,
aps, prl,
floatfix,
twocolumn,
]{revtex4}

\usepackage{graphicx}
\usepackage{bm}
\usepackage{xcolor}
\usepackage{amsmath}
\usepackage{bm}

\begin{document}


\title{Particle dynamics in deposition of porous films with a pulsed radio-frequency atmospheric pressure glow discharge}


\author{Yu Xu}
\affiliation{College of Science, Donghua University, 201620 Shanghai, P. R. China}
\affiliation{State Key Laboratory for Modification of Chemical Fibers and Polymer Materials  and College of Materials Science and Engineering, Donghua University, Shanghai 201620, P. R. China}
\author{Sergey A. Khrapak}
\affiliation{Institut f\"ur Materialphysik im Weltraum, Deutsches Zentrum f\"ur Luft- und Raumfahrt (DLR), 82234 We{\ss}ling, Germany}
\author{Ke Ding}
\affiliation{College of Science, Donghua University, 201620 Shanghai, P. R. China}
\author{Mierk Schwabe}
\affiliation{Institut f\"ur Materialphysik im Weltraum, Deutsches Zentrum f\"ur Luft- und Raumfahrt (DLR), 82234 We{\ss}ling, Germany}
\author{Jian-Jun Shi}
\author{Jing Zhang}
\affiliation{State Key Laboratory for Modification of Chemical Fibers and Polymer Materials  and College of Materials Science and Engineering, Donghua University, Shanghai 201620, P. R. China}
\author{Cheng-Ran Du}
\email{chengran.du@dhu.edu.cn}
\affiliation{College of Science, Donghua University, 201620 Shanghai, P. R. China}


\date{\today}

\begin{abstract}
Nanoparticles grown in a plasma are used to visualize the process of film deposition in a pulsed radio-frequency (rf) atmospheric pressure glow discharge. Modulating the plasma makes it possible to successfully prepare porous TiO$_2$ films.  We study the trapping of the  particles in the sheath during the plasma-on phase and compare it with numerical simulations. During the plasma-off phase, the particles are driven to the substrate by the electric field generated by residual ions, leading to the formation of porous TiO$_2$ film. Using video microscopy, the collective dynamics of particles in the whole process is revealed at the most fundamental ``kinetic'' level.
\end{abstract}


\maketitle


Porous films deposited on substances have many applications such as perovskite-sensitized \cite{burschka2013nature} and dye-sensitized \cite{chen2009advancedmaterials} solar cells, Li-ion batteries \cite{qin2018ami}, sensing \cite{kimura2012advancedfunction,liu2013SensorsandActuatorsB}, and photocatalysis \cite{Varshney2016ChemistryReviews}. The recent development of atmospheric pressure glow discharges (APGD) makes the process of plasma assisted film deposition easier than traditional methods \cite{chen2009advancedmaterials,mocvd2007am} since it can be performed in an open system with high growth rates at low thermal budgets \cite{boysen2018acie,Massines2012ppp,JJShi2006prl}. However, the films deposited by plasma enhanced chemical vapor deposition (PECVD) are usually dense \cite{Hodgkinson2013surfacecoating,li2019ass,Kamal2017acs}. It is challenging to deposit mesoporous films with particles in plasmas on the substrate since the particles are charged by the electron and ion currents and confined by the plasma potential \cite{Selwyn1989jvst,Selwyn1990apl}, preventing them from being deposited \cite{Bazinette2016ppp}.

The dynamics of mesoscopic particles in plasmas has been widely studied in dusty plasma research \cite{Fortov2005physicsreport,shukla2002book,bouchoule1999book}. Using video microscopy, the particle motion can be easily recorded in real time. In a low pressure plasma, the dynamics of a strongly coupled dusty plasma are virtually undamped \cite{Morfill2009RevModPhys,Chaudhuri2011SoftMatter,kryuchkov2018SoftMatter}. This makes dusty plasmas an ideal model system to study regular liquids and solids at the kinetic level \cite{Thomas1996nature,Zuzic2006nature,Tsai2016nature,Wong2018nature,kryuchkov2018prl}.  However, in atmospheric plasmas, the collective dynamics of charged particles is fully damped, and has been barely studied.

In this work we use nanoparticles grown in a plasma to visualize the process of film deposition and demonstrate that modulating the plasma makes it possible to successfully prepare porous TiO$_2$ films. The collective dynamics of the particles driven by the electric field in the initial stage of the plasma-off phase is revealed using video microscopy. We show that varying the plasma-off time allows to control the size of the deposited particles.

\begin{figure}
  \centering
  \includegraphics[width=8.5 cm]{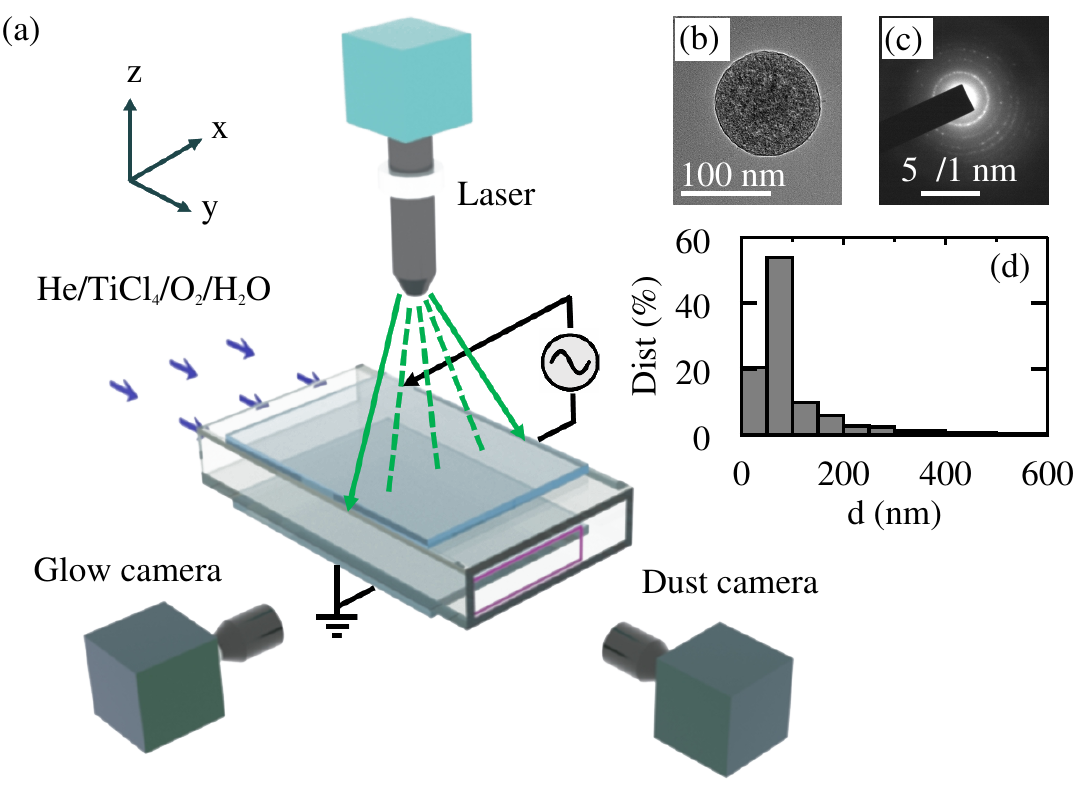}
  \caption{\label{setup} (color online) Illustration of the experiment setup (a). Two cameras and a laser diode are mounted on a translation stage, which can scan the particle cloud in $y$ direction. Morphology (b), SAED pattern (c), and size distribution (d) of TiO$_2$ particles formed in the plasma and collected at the gas exit of the chamber are characterized by TEM. }
\end{figure}

The experiments were conducted in a radio-frequency (13.56~MHz) dielectric barrier discharge system at atmospheric pressure, as shown in Fig.~\ref{setup}(a). A plasma was ignited in a capacitively coupled quartz discharge chamber. The discharge gap was $2$~mm high and $10$~mm wide. Two fluorine-doped tin oxide glass plates were used as electrodes, where the upper one was driven and the lower one was grounded. The particles were illuminated by a laser with wavelength of $532$~nm \cite{thomas1999pop,thomas2004pop,schlebrowski2013psst,barbosa2015jpd,du2017njp}. Two CMOS cameras were used to record the particle dynamics and the evolution of the discharge glow with a frame rate of $100$ frames per second (fps). Helium was used as discharge carrier gas for TiCl$_4$ and H$_2$O, and O$_2$ was used as the oxidizing agent. The corresponding gas flow rates were set at $500$, $10$, $20$ and $5$~sccm, respectively.

The deposition of a porous film was achieved by pulse modulating the rf discharge. After igniting the helium  plasma,  we inlet carrier gases into the discharge chamber. The input voltage was set to $U_{pp}\sim2500$~V. Since the plasma conditions play an important role in the particle formation, we kept the plasma-on time $t_{on}=100$~ms, and varied the time of the plasma-off phase for $t_{off}=300$, $200$, $100$, $50$, $25$~ms, corresponding to the repetition frequencies  $f=1/(t_{on}+t_{off})=2.5$, $3.3$, $5$, $6.6$, and $8$~Hz. The particles grew in the plasma.

To characterize the TiO$_2$ particles formed in the plasma, we collected the dust particles at the gas exit of the discharge chamber and performed transmission electron microscope (TEM) analysis to determine their sizes and structure. As shown in Fig.~\ref{setup}(b), the particles have a spherical shape.  In the Selected Area Electron Diffraction (SAED) pattern in Fig.~\ref{setup}(c), the polymorphic rings indicate that the TiO$_2$ particles are polycrystalline. The distribution of particle diameters was measured from the TEM images and shown in Fig.~\ref{setup}(d). The average diameter of the collected particles is $98$~nm. Larger particles can be formed in the plasma discharge but their amount is rather limited.

The polycrystalline TiO$_2$ particles are formed  in the He/TiCl$_4$/O$_2$/H$_2$O atmospheric pressure plasma. The precursor molecules are excited/dissociated/ionized into active precursor species by the plasma first. Then they react and nucleate into nanoscale particles, agglomerate and further grow into submicron particles as displayed in Fig.~\ref{setup} \cite{bouchoule1999book,mitic2018ppp,mikikian2016ppcf,berndt2009cpp,couedel2010pop}. This process is accompanied by the formation of negatively charged clusters and the agglomeration of these clusters when critical densities are reached; further growth by  radical sticking  is halted when charging effects prevent further agglomeration. In our experiment with the continuous discharge, no TiO$_2$ film can be deposited inside the discharge zone. This means that negatively charged TiO$_2$ particles are formed as the precursor molecules enter the plasma discharge zone and become trapped near the sheath. The film deposition marginally depends on the diffusion and surface reaction of
active species, but is achieved with a different process.

\begin{figure}
  \centering
  \includegraphics[width=8.5 cm]{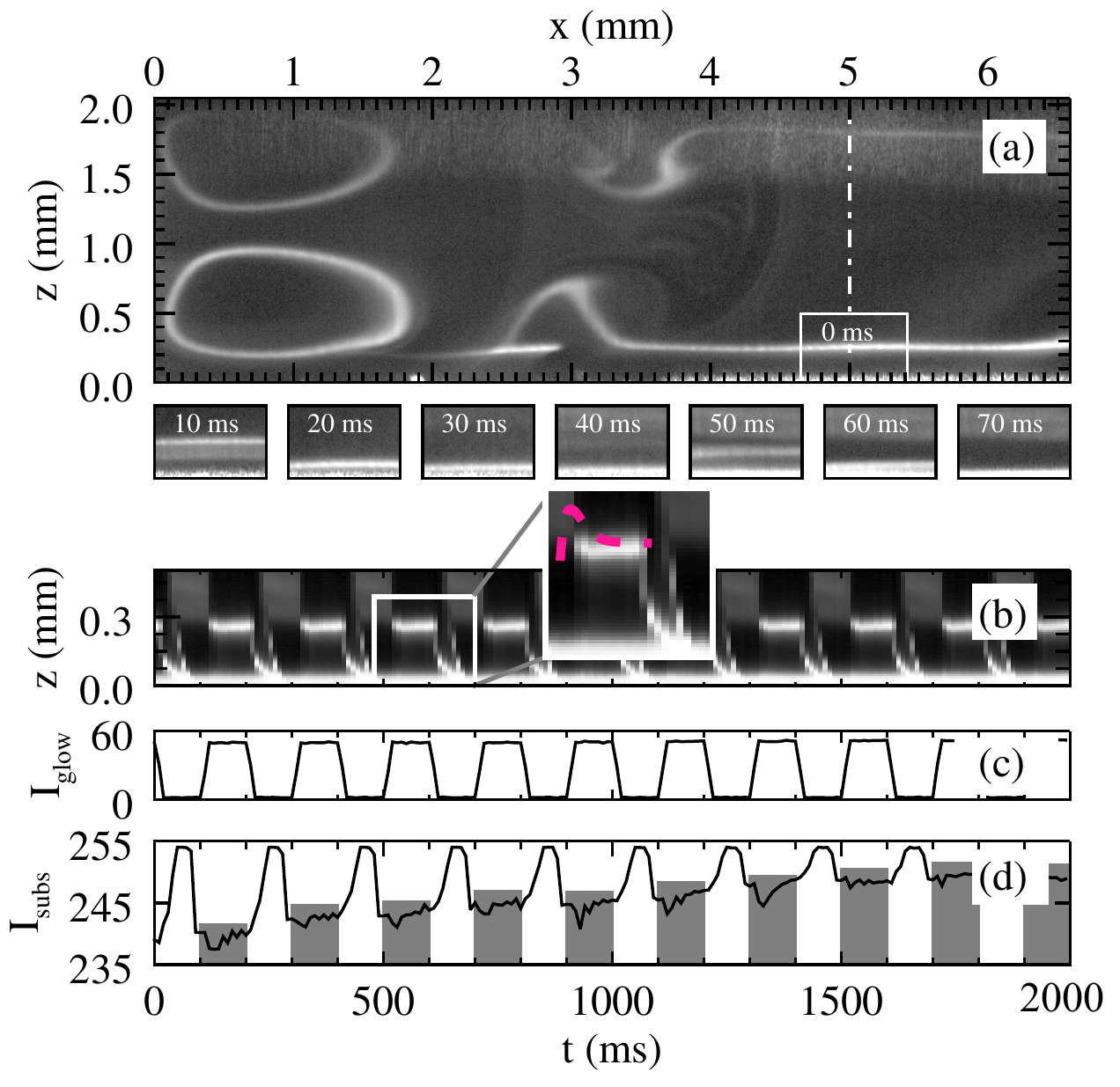}
  \caption{\label{dynamics} (color online) Particle dynamics and plasma glow for $t_{off}=100$~ms. The spatial distribution of TiO$_2$ particles during the plasma-on phase is shown in (a) (time is set as $0$~ms  one frame before the plasma-off phase starts). The vertical dashed line marks the symmetry axis of the chamber. The ROI is highlighted by the white rectangle. The profiles of the particle cloud in the ROI for the next $7$ consecutive frames are shown in a row beneath, for a time period of $70$~ms. The spatiotemporal evolution of the particle cloud for ten periods is shown in (b), and the sheath edge found in the simulation is illustrated by the red dotted curve overlaid in the magnified inset. The glow intensity is plotted in (c) and the intensity of the laser light reflected by the substrate is shown in (d).}
\end{figure}

During the plasma-on phase of the modulated discharge, the negatively charged particles close to the axis of symmetry [dashed-dotted line in Fig.~\ref{dynamics}(a)] are confined in two layers close to the bottom and top electrodes with a huge void in-between \cite{Mikikian2010prl}, as shown in Fig.~\ref{dynamics}(a). The ion drag force \cite{kharapakpre2002} is balanced by the electric force in the (pre)sheath, and gravity plays a minor role. Close to the side of the chamber, vortexes are clearly visible, presumably caused by the thermophoretic force and neutral drag \cite{Schwabe2014prl,Zuzic2007njp}. Between the central stable region and the vortexes, there appear bent layers of particles at the position $x\approx3$~mm. The detailed origin of theses vortexes and bent layers is beyond the scope of this manuscript and will be described elsewhere. In this paper, we focus on the dynamics of particles in the region of interest (ROI) close to the chamber center, marked by a white rectangle in Fig.~\ref{dynamics}(a).

As the plasma-off phase starts, the glow disappears instantaneously in terms of the recording rate, as shown in Fig.~\ref{dynamics}(c). The particles can no longer be confined in the (pre)sheath. We show the particle positions in the ROI in $8$ consecutive frames recorded by the ``dust camera'' with a time interval of $10$~ms in Fig.~\ref{dynamics}(a). As we can see, the cloud drops to the bottom electrode within $30$~ms followed by a second drop at $50$~ms. This dynamics can also be clearly seen in the spatiotemporal evolution in Fig.~\ref{dynamics}(b). Particularly, the particle layer splits into two layers (at $10$~ms) as it falls down. The process of the (visible) particle deposition lasts for $t_c\approx70$~ms. After $100$~ms, the plasma is again ignited. The intensity of the glow, recorded by the ``glow camera'', restores to a value of $60$. The particles are trapped in the (pre)sheath again. However, the height of the particle layer shows a spike as soon as the plasma is ignited, and it restores to its equilibrium position afterwards. It is well known that dust particles can be used as  diagnostics for the sheath  \cite{Annaratone2004prl,Annaratone2003njp,Kersten2003njp,Basner2009njp}. This result implies that a wider sheath is formed at the moment of ignition.

We performed a dedicated numerical simulation to study the evolution of the plasma sheath as the voltage is applied. The model simulates an atmospheric pressure, rf-driven capacitive discharge using a hybrid analytical-numerical global model. The model's detailed description is given in the supplemental material as well as in Ref. \cite{keding2014psst,ding2014jpd,lazzaroni2012psst}. In the simulation,  the sheath width increases to $0.33$~mm after the ignition of plasma. It decreases to a stable value of $0.26$~mm afterwards, as shown by a pink-dashed curve in the inset of Fig.~\ref{dynamics}(b). The spike of the height of the particle layer observed in the experiment is much smaller than the spike of the sheath width in the simulation. This is caused by the friction of particles in the atmospheric plasma, where the damping rate is about $\nu_n \simeq 10^6$~s$^{-1}$. The particles cannot follow the formation of the sheath. Nevertheless, the experimental results show a qualitative agreement with the numerical simulation.

The particle acceleration towards the electrode in the plasma-off phase can be explained as follows. When the plasma is on, the particles are charged negatively due to the high electron thermal velocity. The estimated average particle charge number is $Z\simeq 100$ \cite{Khrapak2006pop,Khrapak2009cpp}.  After the plasma is off, the electron population is expected to vanish on a time scale of $10^{-5}$ s, while the ion density remains almost unchanged. The charge is now determined by the balance of positive and negative ion fluxes. Due to the higher mobility of positive ions (O$_2^+$) than the negative ions (Cl$^-$) in He gas \cite{Ellis1976}, the particle charge will shift to the positive values ($Z\simeq 0.8$). Most of the particles are either uncharged or carry one or two elementary charges.

When the electron component has just vanished, the electric field (resulting from the space charge of ions) would accelerate a singly charged particle up to a terminal velocity $v_p$, determined from the balance between the electric force and friction against the neutral gas, $v_p\simeq eE/m_p\nu_n\sim 100$ cm/s, where $e$ refers to the electron charge, $E$ denotes the electric field, $m_p$ denotes the particle mass, and $\nu_n$ denotes the frictional damping rate. The electric field decays as the ions start to diffuse towards the discharge walls and electrodes. When  most of the ions are lost (on the time scale of few ms), all particle charges remain ``frozen'' since no mechanism of re-charging is available. Meanwhile, the electric field should not necessarily vanish, because some positive charge can be carried by a fog of small nanometer size particles due to the same frozen-charge effect. The remaining electric field pushes discretely charged particles ($1$~e or $2$~e) to the electrode, resulting in the two dropping layers of particles at $t\sim10$~ms, which agrees with experimental observations. The details of these estimations are provided in the supplemental materials.

\begin{figure}
  \centering
  \includegraphics[width=8.5 cm]{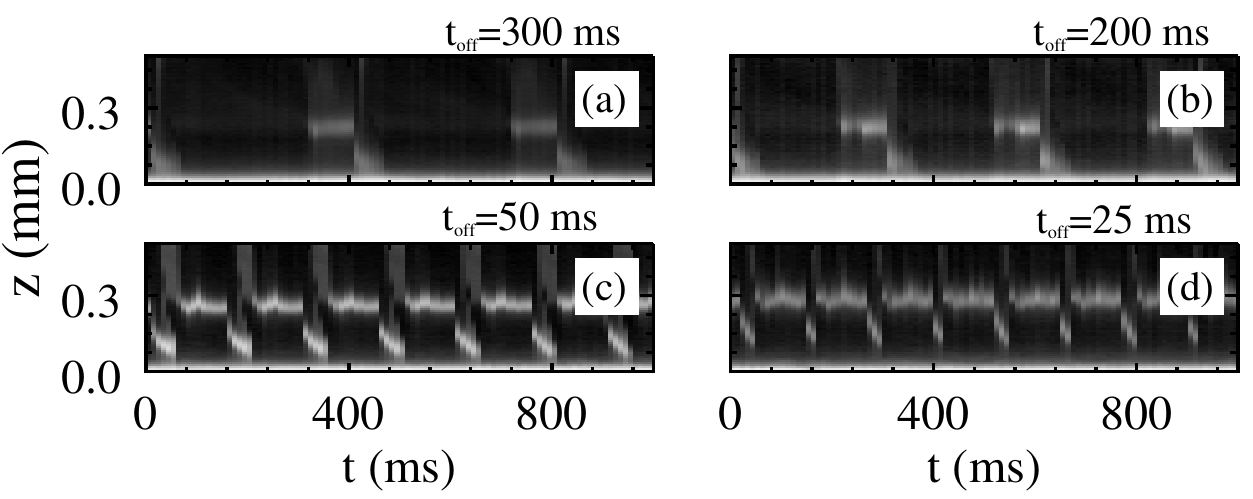}
  \caption{\label{4 dynamics} Spatiotemporal evolution of the particle cloud for $t_{off}=300$~ms (a), $200$~ms (b), $50$~ms (c), and $25$~ms (d).}
\end{figure}

As TiO$_2$ particles are accelerated to the substrate by the modulated rf discharge, a porous film is deposited. The reflection of laser light on the substrate can indicate the thickness of the film. As we can see in Fig.~\ref{dynamics}(d), the intensity of the reflection on the substrate increases stepwise as particles are accelerated to the substrate periodically \footnote{The intensity of the reflection has a non-linear relation to the thickness of the film. But generally, the brighter the reflection is, the thicker the film is.}.

Since the modulation plays an important role in film deposition, we study the particle dynamics with various plasma-off times. The results are shown in Fig.~\ref{4 dynamics}.  For $t_{off}=200$ and $300$~ms, the plasma-off phase is long enough that almost all dust particles can reach the substrate. However, when the time of the plasma-off phase is shorter than the critical time $t_c$($\approx 70$~ms), the next discharge ignites before most particles reach the substrate. The particles are moved back to the sheath edge, and thus the deposition process is inhibited by the fast formation of the sheath in the plasma.

\begin{figure}
	\centering
	\includegraphics[width=8.5 cm]{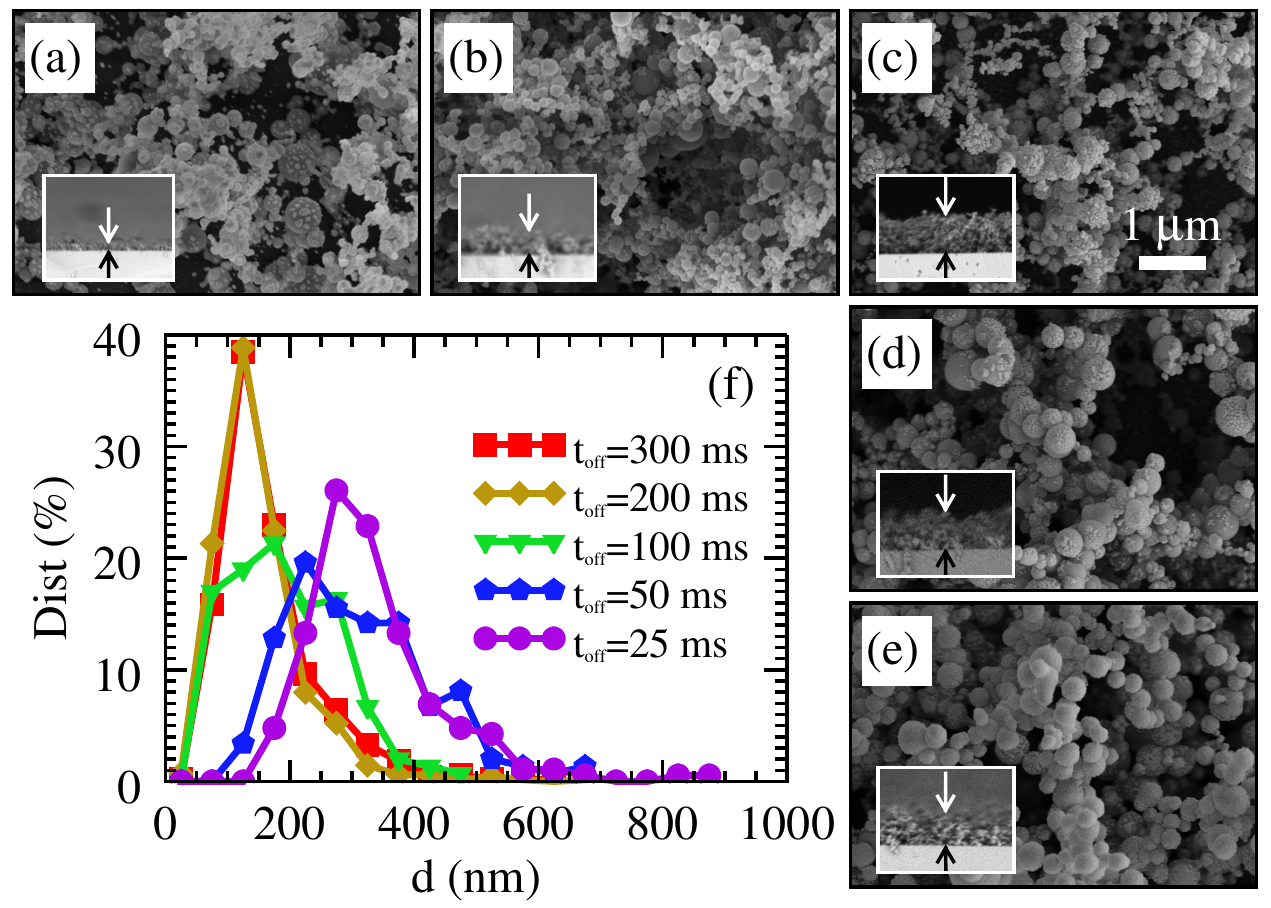}
	\caption{\label{SEM} (color online) SEM images of the deposited porous film for $t_{off}=300$~ms (a), $200$~ms (b), $100$~ms (c) $50$~ms (d), and $25$~ms (e) and size distribution of particles as constituents of the film (f). The insets show the side view of the corresponding films with a total inset height of $40$~$\mu$m. }
\end{figure}

We study the morphology of the TiO$_2$ films deposited for $5$ minutes under the scanning electron microscope (SEM). Images of the surface and the cross section of the films deposited with five plasma-off times are shown in Fig.~\ref{SEM}(a-e). For $t_{off}=200$ and $300$~ms, the surface morphology of the film is rather similar, as shown in Fig.~\ref{SEM}(a,b). As  $t_{off}$ decreases, the ratio of big particles to small particles increases. For $t_{off}=25$~ms, only big particles can be seen as constituents of the film, as shown in Fig.~\ref{SEM}(e). This trend is clearly demonstrated in the distribution of particle diameters shown in  Fig.~\ref{SEM}(f). The peak lies at about $100$~nm for $t_{off}=200$ and $300$~ms, similar to the size distribution of the particles collected at the gas exit, as shown in Fig.~\ref{setup}(d). However, the peak shifts to larger diameter and becomes broader as $t_{off}$ decreases. Here only the bigger particles (with higher probability of being charged) get accelerated enough that they can reach the substrate within the limited time of the plasma-off phase. Most particles, in particular those with small size, are turned back to the sheath as the plasma is ignited after the very short off time.

\begin{figure}
  \centering
  \includegraphics[width=8.5 cm]{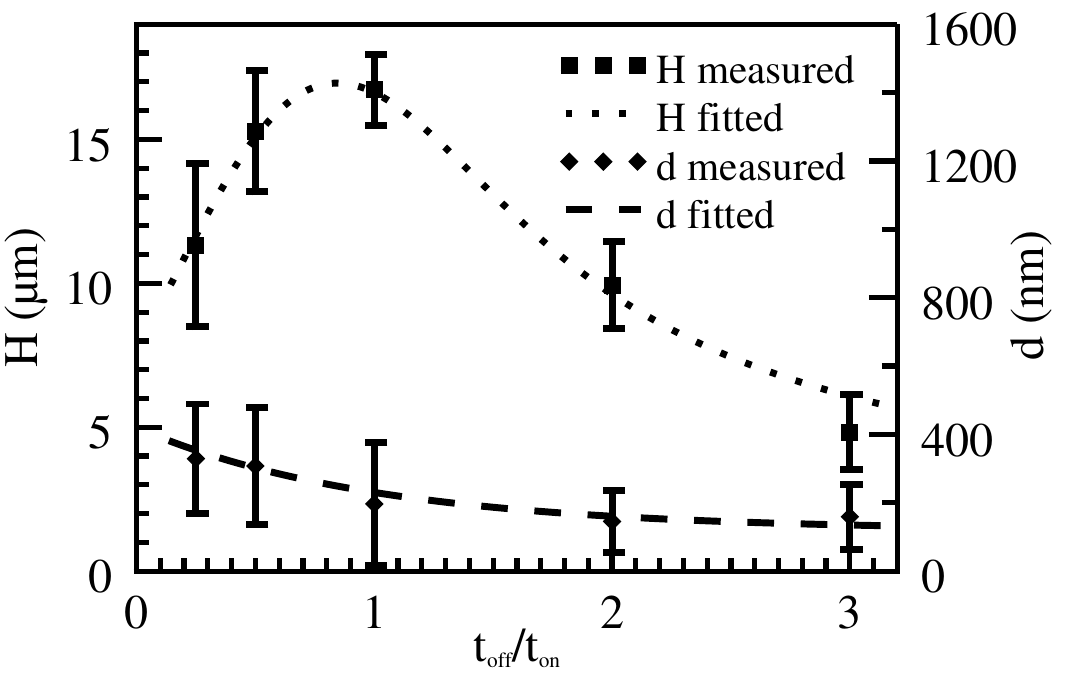}
  \caption{\label{thickness} Dependence of the film thickness $H$ and mean diameter of deposited particles $d$ on the ratio of plasma-off time $t_{off}$ to plasma-on time $t_{on}$. }
\end{figure}

As a consequence of the particle dynamics, the thickness of the film varies considerably with the plasma-off time, as shown in the insets of Fig.~\ref{SEM}. Here we measure the TiO$_2$ film thicknesses and mean particle diameters with a total deposition time of $5$ minutes. As we can see in Fig.~\ref{thickness}, the film thickness first increases from $12$~$\mu$m to $18$~$\mu$m as $t_{off}$ increases up to $100$~ms. It then decreases to $6$~$\mu$m for $t_{off}=300$~ms. Two factors influence the film thickness, i.e., the repetition frequency and the duration of the plasma-off phase. On the one hand, a longer plasma-off time leads to fewer repetitions within a certain time period. For  $t_{off}=100$, $200$, $300$~ms, almost all particles can be deposited on the substrate within each repetition. Therefore, the more repetitions there are, the thicker the film is. On the other hand, a shorter plasma-off phases prevents small particles from reaching the substrate. Consequently, fewer particles are deposited on the substrate in each repetition. This causes the thinner film as $t_{off}$ further decreases below $100$~ms. As to the size of the particles in the porous film, the mean diameter keeps relatively stable for $t_{off}>100$~ms and increases as a shorter plasma-off time is applied, as shown by the diamond symbols in Fig.~\ref{thickness}. Based on these factors, we can describe the dependence of the film thickness $H$ on the plasma-off time $t_{off}$ with a phenomenological formula:
\begin{equation}
H=(\frac{A}{R_t+1})\left(Be^{-R_t}+C\right)\left\{\frac{1}{2}\left[1+erf\left(DR_t-E\right)\right]\right\},
\label{eq_thickness}
\end{equation}
where $R_t(=t_{off}/t_{on})$ is the ratio of the plasma-off time to the plasma-on time. The formula is composed of three terms, where the first term, $A/(R_t+1)$, is associated with the repetition frequency, the second term, $Be^{-R_t}+C$, represents the effect of particle sizes, and the third term, $1/2[1+erf(DR_t-E)]$, is the cumulative (normal) distribution function \footnote{The cumulative distribution function should approach one as $t_{off}$ exceeds the critical time $t_c$}. The formula (\ref{eq_thickness}) is fitted to the measured film thickness $H$ and its second term is fitted to the mean particle diameter $d$, resulting in the fitting parameters as $A=90$, $B=0.3$~$\mu$m, $C=0.12$~$\mu$m, $D=1.5$, and $E=0.9$. As we can see in Fig.~\ref{thickness}, this formula provides a good agreement to the experimental results.

In summary, we apply a pulse modulation to overcome the particle trapping effect and successfully prepare porous TiO$_2$ films in an rf atmospheric plasma glow discharge. In contrast to film deposition by plasma-jet \cite{Homola2016acs}, the thickness and the structure of the porous films can be controlled directly by tuning the discharge parameters such as the plasma-off time. Using video microscopy, we directly observe the collective dynamics of particles during the deposition, revealing the whole process at the most fundamental ``kinetic'' level. Besides, the formation of the plasma sheath at the ignition is monitored using particles as diagnostics, showing a fair agreement with the numerical simulations. This work sheds light on the microscopic mechanism of the plasma assisted porous film deposition and extends its application to various fields requiring fine control.

\begin{acknowledgments}
The authors acknowledge support from the National Natural Science Foundation of China (NSFC), Grant No. 11405030, 10775031, and 10835004. We thank Michael A. Lieberman and Hubertus M. Thomas for the valuable suggestions.
\end{acknowledgments}

\end{document}